\documentclass{elsart3}
\usepackage{graphicx}
\usepackage{subfigure}
\usepackage{latexsym}

\begin{document}

\newcommand{\mrm}{\mathrm}

%%upright Greek letters (example below: upright "mu")
\newcommand{\greeksym}[1]{{\usefont{U}{psy}{m}{n}#1}}
\newcommand{\umu}{\mbox{\greeksym{m}}}
\newcommand{\udelta}{\mbox{\greeksym{d}}}
\newcommand{\uDelta}{\mbox{\greeksym{D}}}
\newcommand{\uOmega}{\mbox{\greeksym{W}}}
\newcommand{\uPi}{\mbox{\greeksym{P}}}
\newcommand{\ualpha}{\mbox{\greeksym{a}}}

\sloppy
\begin{frontmatter}
\title{Development of an Indium Bump Bond Process for Silicon Pixel Detectors at PSI}
\author[psi]{Ch.~Broennimann}
\author[psi]{F.~Glaus}
\author[psi]{J.~Gobrecht}
\author[psi]{S.~Heising}
%\author[psi]{B.~Henrich}
\author[psi]{M.~Horisberger}
\author[psi]{R.~Horisberger}
\author[psi]{H.~C.~K\"astli}
\author[psi]{J.~Lehmann}
\author[psi]{T.\,Rohe\thanksref{corr}}
\author[eth]{S.~Streuli}
\address[psi]{Paul Scherrer Institut, 5232 Villigen PSI, Switzerland}
\address[eth]{ETH~Z\"urich, IPP, 5232 Villigen PSI, Switzerland}

\thanks[corr]{Corresponding author; e-mail: Tilman.Rohe@psi.ch}

\begin{abstract}
The hybrid pixel detectors used in the high energy physics experiments
currently under construction use a vertical connection technique,
the so-called bump bonding. As the pitch below $100\,\umu$m, required in these 
applications, cannot be fullfilled with standard industrial processes 
(e.g. the IBM C4 process), an in-house bump bond process using reflowed
indium bumps was developed at PSI as part of the R\&D for the CMS-pixel detector.

The bump deposition on the sensor is performed in two subsequent lift-off steps.
As the first photolithographic step a thin under bump metalization (UBM) is 
sputtered onto bump pads.
It is wettable by indium and defines the diameter of the bump. The indium is 
evaporated via a second photolithographic step with larger openings and is 
reflowed afterwards. The height of the balls is defined by the volume of the indium. 
On the readout chip only one photolithographic step is carried out to
deposit the UBM and a thin indium layer for better adhesion. After mating
both parts a second reflow is performed for self alignment and obtaining
high mechanical strength.

For the placement of the chips a manual and an automatic 
machine were constructed. The former is very flexible in handling
different chip and module geometries but has a limited throughput while
the latter features a much higher grade of automatisation and is therefore
much more suited for producing hundreds of modules
with a well defined geometry.

The reliability of this process was proven by the successful construction of
the PILATUS detector. The construction of PILATUS~6M (60 modules) and the
CMS pixel barrel (roughly 800 modules) will start in 2005.
\end{abstract}
\end{frontmatter}

\section{Introduction}
The CMS experiment, currently under construction
at the Large Hadron Collider (LHC) at CERN (Geneva, Switzerland),
will contain a hybrid pixel detector for tracking and vertexing \cite{cms-pixel}.
It requires bump bonding with a minimal pitch of $100\,\umu$m which is 
below the industrial standard. In order to achieve the highest
flexibility during prototyping and a fast turn-over an in-house bump bond 
process was developed. It makes use of the infrastructure present at
the Paul Scherrer Institut (PSI). The process was successfully applied to the 
PILATUS~1M detector \cite{gregor}, a single photon counting hybrid pixel 
detector with an area of $24.3\times 20\,$cm$^{2}$.

\begin{figure}
\centering\includegraphics[width=1\linewidth]{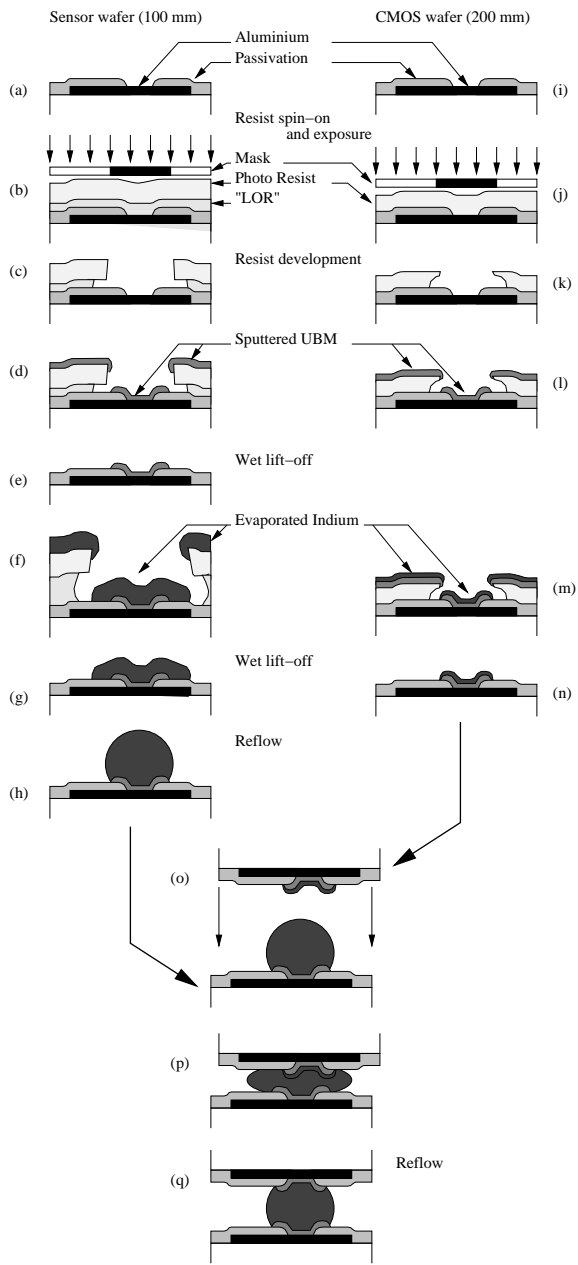}
\caption{Schematic flow diagram of the bump bonding process consisting of the
	bump deposition onto the sensor (a)-(h), metal deposition on
	the readout chip (i)-(n), and the flip chip procedure (o)-(q).
	\label{fig:process}}
\end{figure}

In this process sketched in Fig.~\ref{fig:process} the bumps are deposited 
onto the sensor part of the hybrid pixel module while only a thin 
($\approx 1\,\umu$m) indium layer is deposited just to increase the 
adhesion. The bumps (deposited on the sensor)
are reflowed to balls before the parts are joined. For self alignment and
to provide good mechanical strength a second reflow is performed after
the chip flip procedure.

\section{Bump deposition on the sensor}

The metal deposition on the sensor is done in two
separate lift-off steps followed by a reflow.

\subsection{Underbump metalization}

The under bump metalization (UBM) consists of three
metal layers which are added on top of the last metalization
layer of the sensor (see Fig.~\ref{fig:process}~(a)-(g)). The first 
one is about 10\,nm titanium that acts as a barrier and an adhesive 
layer. It is followed by roughly 50\,nm nickel which is wettable 
with indium and a roughly 50\,nm thick protective gold layer. 

This sandwich is deposited in a so-called lift-off process. 
First (after a cleaning procedure) a two-layer photoresist is spun 
onto the wafer. The lower one is a $1.2-1.5\,\umu$m thick lift-off resist (LOR) which
is not light sensitive. The top layer is a negative resist with 
a thickness of about $3.5\,\umu$m. 
{\em Negative} means that the resist gets developed in the
non-exposed areas and a light-field mask is
used on which the bumps appear as dark points. 
After exposure (Fig.~\ref{fig:process}~(b)) the non-exposed
parts of the resist are developed. As the LOR is not light
sensitive it is over-developed and an overhanging edge, a so
called {\em undercut}, is created (Fig.~\ref{fig:process}~(c)).

The choice of the resist polarity is purely historical. In the first attempts
``generic'' masks were used and the alignment was done using the
bumps themselves. This requires a mask which is essentially
transparent. The disadvantage of negative resist is its sensitivity
to stray light. If during exposure light is scattered at a rough
metal surface into the shadowed region of the bumps, some resist
might harden there. This leads to an undeveloped 
resist layer remaining on the bottom of the openings. The deposited 
metal layer is removed with the resist. As the sensor
metalization is large around the bump pad and the surface 
is rather rough light scattering appears sometimes, but this
effect is eliminated by the use of the LOR. 

The deposition of the three metal layers (Fig.~\ref{fig:process}~(d))
is done in a multi-target magnetron sputtering machine. The thickness of the 
UBM is very small and cannot cover the overhanging edges of the resist.

When the photo resist is removed (lifted-off), the thin metal foil on 
top of it is also removed. The metal remains only in the area of the
bump pad (Fig.~\ref{fig:process}~(e)).
As the LOR is not soluble in acetone, this process has to be
done with a special remover %(pH-value close to 7) 
at an elevated temperature of about
$60\,^\circ\,$C. The area of the deposited metals is very small causing this lift-off
process to be slow. Process times of more than 12~hours are common.

\subsection{Indium bump deposition}

The indium bumps are deposited using a similar lift-off process. The only
difference is the thickness of the resist and the increased size of the openings. 
To safely exceed the 
layer thickness of the evaporated indium of $2-3\,\umu$m, the resist
is spun on with a total thickness of about $8\,\umu$m ($3.5\,\umu$m LOR plus
$4.5\,\umu$m photoresist).

\begin{figure}
\centering\includegraphics[width=0.5\linewidth]{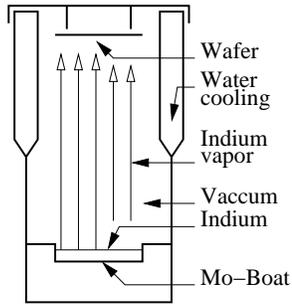}
\caption{Schematic cross section through the indium evaporation 
	vessel. When the vacuum is established the indium in the
	molybdenum boat is heated. It evaporates and condenses
	on the wafer mounted on top.
	\label{fig:in-vessel}}
\end{figure}
The indium is deposited in a 50\,cm high cylindrical vessel (see 
Fig.~\ref{fig:in-vessel}) with a 
diameter of about 30\,cm that can be evacuated down to a few $10^{-5}$\,mbar.
When  this level is reached the indium in a molybdenum
boat at the bottom is heated. It evaporates and condenses on the wafer
mounted at the removable top cover. The side wall of the vessel is water-cooled  
which considerably reduces the thermal budget on the photoresist. By this measure
blaining of the resist is prevented. Further the resist hardens if it is 
exposed to too much heat and its removal becomes difficult.

The removal of the photoresist is done similarly to the UBM process.
After the lift-off the indium forms a rather flat octagon as
shown in Fig.~\ref{fig:bump-foto}~(a). As the bumps are
relatively insensitive to mechanical damage in this state and 
the wafers can be handled without special care, it is the right 
moment to dice the wafers and again test the individual sensors
on a probe station.

\subsection{Reflow\label{sec:reflow-1}}

\begin{figure*}[t]
\subfigure[]{\includegraphics[width=0.48\linewidth]{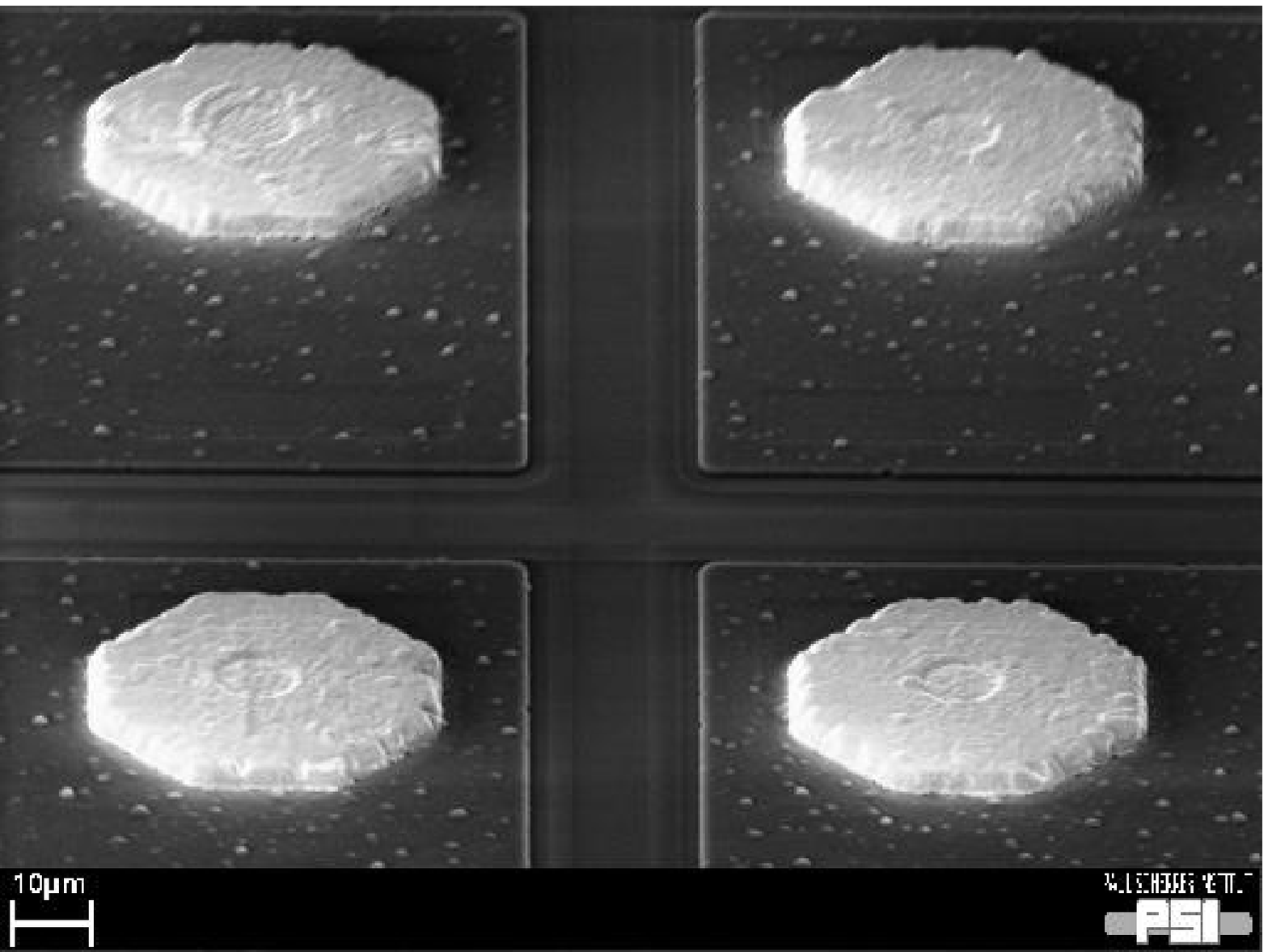}}
\hfill
\subfigure[]{\includegraphics[width=0.48\linewidth]{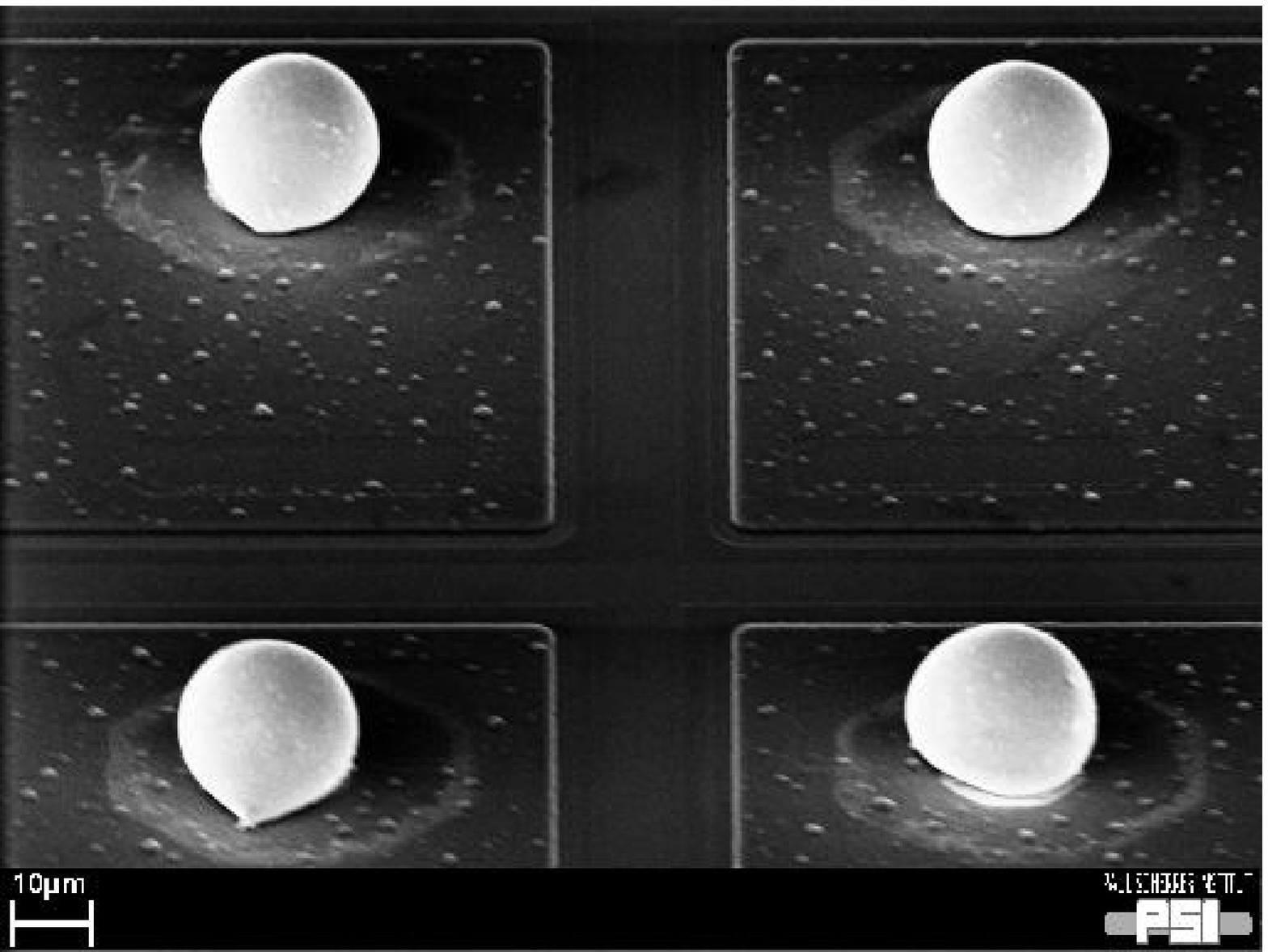}}
\caption{A scanning electron micrograph of indium bumps before (a) and 
	after (b) reflow. The distance between the bumps is $100\,\umu$m,
	the deposited indium is $50\,\umu$m wide, the
	reflowed bumps have a diameter of about $20\,\umu$m.\label{fig:bump-foto}}
\end{figure*}

In order to form balls from the indium the devices have to be
heated. When the indium is molten, the balls are created by the surface tension.
The size of the ball is defined by the diameter of the wettable UBM pad and
the volume of the evaporated indium. The result of the reflow process is shown 
in Fig.~\ref{fig:bump-foto}. The distance between the bumps is $100\,\umu$m. 
The width of the deposited indium octagon is in this case $50\,\umu$m and its 
thickness is about $2\,\umu$m. This leads to a volume of about $4000\,\umu$m$^3$
per bump or a bump diameter of about $20\,\umu$m.

The reflow is performed in a microprocessor controlled oven, regulating
temperature profile, flux gas pressure, etc. which was designed and built at PSI.

\section{Metal deposition on the readout chip}

For the readout chips the same lift-off mask is used for the 
UBM and the indium as indicated in Fig.~\ref{fig:process}~(i)-(n). 
This is possible because the amount of indium deposited on the
chips is much smaller and no bump balls will be formed here.
As the metal surface of the CMOS wafers is usually very 
plain and stray light is not
an issue, the use of lift-off resist (LOR) is not necessary. Only
a layer of about $3.5\,\umu$m negative resist is spun onto the
200-mm-wafers. The undercut of the edges is created by
a suitable choice of exposure and development times. 

The UBM is then sputtered onto the wafer directly followed by
the evaporation of a thin $1-2\,\umu$m indium layer.
Due to the use of ``normal'' photoresist, the lift-off can be
done at room temperature using acetone.

After the lift-off process the wafers are thinned and diced.
The task of the thin indium layer is to improve adhesion 
after the chip flipping. The chips are however not
reflowed before the flip chip procedure.

\section{Chip placement}

The readout chips and the sensors are joined to form a 
{\em bare module}. As in most cases several
read-out chips (16 for CMS and PILATUS modules) are placed 
onto one sensor, the latter is mounted on a table and the chips are 
placed successively. This is done either in a
manual or automatic chip placement machine. While the former
is very flexible and used for prototype and single chip
assemblies, the latter is automated to a high degree and 
allows the throughput necessary for the construction of 
larger pixel detectors. After the placement of the chips, 
the modules are reflowed to establish the mechanical connection. 
Further the surface tension of the molten indium provides 
self-alignmet (see Fig.~\ref{fig:process}(o)-(q)).

After the reflow, the module is robust enough for further 
handling and will be tested before being fed it into the
module production line.

\subsection{Manual chip placement}

\begin{figure}
\centering\includegraphics[width=1\linewidth]{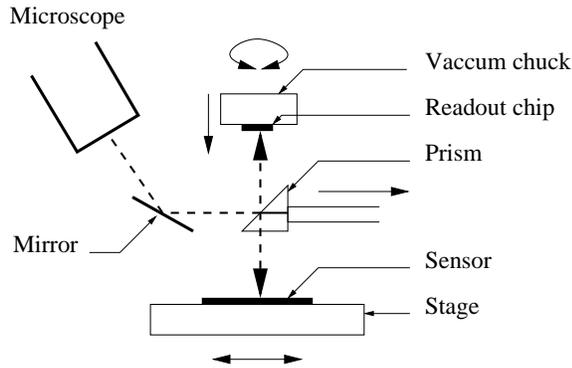}
\caption{Schematic diagram of the working principle of the manual chip 	
	placement machine. Both parts are aligned in respect to each
	other using a prism placed in between them. After alignment
	the prism is removed and the chip is pressed onto the
	sensor. \label{fig:manual}}
\end{figure}

The manual chip placement machine was designed for the
fast and flexible production of a small quantity of modules
in the prototyping stage of the CMS pixel detector. Its
working principle is schematically shown in Fig.~\ref{fig:manual}.
The sensor is mounted onto a movable stage while the readout chip is
held with its face down on a vacuum chuck. Both parts can 
simultaneously be observed through a prism. When both parts are 
aligned with respect to each other, the prism is removed and 
the chuck with the readout chip is pressed onto the sensor. 
A force up to about 50\,N is applied.

The machine is designed and constructed to flexibly handle
different module sizes and geometries which is very important
in the R\&D phase of a project. However, the manual alignment 
procedure is somewhat tedious and results in 3--4\,h needed
for the placement of 16 chips on a sensor. The throughput
of the machine is limited to a few modules per week which is
not sufficient for building detectors with hundreds of modules.

\subsection{Automatic chip placement}

\begin{figure*}
\centering\includegraphics[width=1\textwidth]{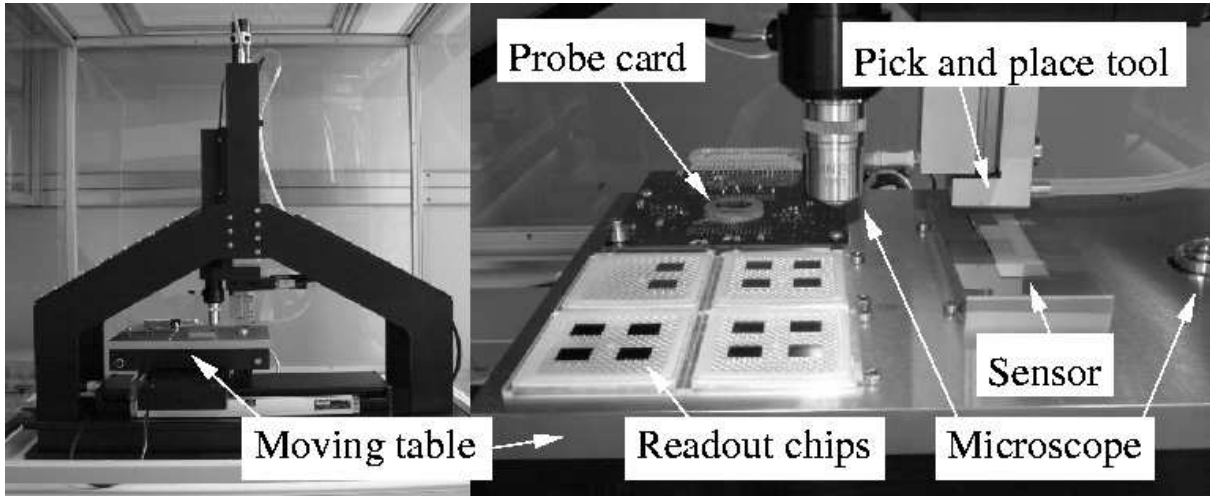}
\caption{Automatic chip placer. \label{fig:auto}}
\end{figure*}

In order to overcome the limitations of the manual chip placement
machine, a fully automated one, shown in Fig.~\ref{fig:auto},
was designed and built. It consists of a table
which can automatically move with a precision of one micrometer 
in the horizontal plane. The vertical movements and possible
rotations are performed with the pick-and-place tool which is fixed to
the very massive bridge.

After equipping the table with the sensor and the readout chips,
the position of the sensor is precisely measured with a microscope
mounted next to the pick-and-place tool. This is done automatically
using alignment marks on the sensor and a pattern recognition algorithm.

Then a first read-out chip placed face down on a gel-pack is picked by the
pick-and-place-tool and held by vacuum. Its position is also precisely
measured with the  microscope looking up which is integrated into
the moving table. The chip undergoes an electrical functionality test using the
probe card mounted on that table. Only working chips are placed on the sensor.
Placing can be done as the sensor and chip position within the machine frame
are known from the position measurements. It is pressed down
with a force of 30~N (about 8-10~mN per bump) and held for 1~min.
This procedure is repeated for all 16~readout chips. 

In case one of the chips does not pass the electrical test, it is put back
onto the gel-pack and a reserve chip from an additional gel-pack (not visible 
in Fig.~\ref{fig:auto}) is used instead. This should rarely happen as all 
chips were tested on-wafer and only good dies are picked from the
dicing tape. However, during the dicing and picking procedure and the subsequent 
handling a small fraction of the chips might be damaged. 

The placement of 16~chips on a sensor takes about 50~min including the
chip test plus about 15~min for loading the machine and preparing the 
components. During the chip placement no human intervention is necessary.
This allows a throughput of several modules per day, sufficient
for the construction of a vertex detector with some hundreds of modules.

\subsection{Reflow}

After the chip placement is finished the assembly undergoes a second
reflow similar to the one described in Sect.~\ref{sec:reflow-1}.
As the module is very fragile before this step, the transfer from
the chip placement machine to the reflow oven has to be done
with special care. After melting a stable mechanical
connection is established. Further the surface tension of the 
indium in addition provides self-alignment.

\subsection{Tests}

The reflowed module, called the {\em bare module}, is robust
enough for handling, and tests can be performed.

\subsubsection{Pull test\label{sec:pull}}

A correctly joined assembly is resistant to about 2~mN pulling force 
per bump which adds up to about 8~N for a CMS-readout chip with 4160 
bump bond connections. To test the quality of the bumps, every
chip is pulled with a force of about 1.8~N applied through a vacuum cup.
If this test fails, which happens very rarely, it is an
indication for a serious mistake either during chip placement or
reflow. In this case the module is excluded from all further 
production steps and the reason for the failure has to be investigated. 

\subsubsection{Geometrical verification\label{sec:geo}}

A very efficient way of testing the correctness of the chip placement is
to measure the distance between readout chips. Due to the self alignment
during the reflow, the distance between the readout chips in a correctly
bump bonded module differs by less than five micrometres from its
nominal value. 

The measurement of the distance can easily be performed using the automatic
chip placement machine. The bare module is positioned with the chips
down on the moving table and the distance between alignment marks on
the chip parts that extend over the edge of the sensor
can be measured automatically.

If a chip is found to be misplaced the module is excluded from all further
processing. This would indicate a serious error during chip placement
or handling. 

\subsubsection{Tests of the readout chips\label{sec:bare-module-test}}

While failure modes described in Sect.~\ref{sec:pull} and~\ref{sec:geo}
are expected to occur only in the prototyping stage, when the procedures
are not yet well established, the experience of other pixel projects 
\cite{alice,alenia,izm} shows that the failure of readout chips 
after bump bonding cannot completely be avoided. The damage is often
caused by silicon pieces which fall off the chip edges during or 
after dicing. If the size of such a piece exceeds the bump height, it is
pressed into the surfaces of the sensor or/and the readout chip which
in most cases causes serious damages such as power shorts in the readout chip.
In order to detect such failures each chip of a bare module is tested
with a probe card. 

\subsection{Rework}

\begin{figure}
\centering\includegraphics[width=1\linewidth]{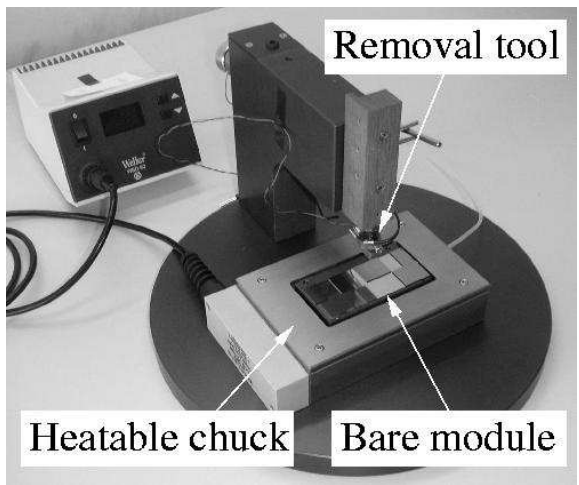}
\caption{Tool for removing a readout chip from a bare module\label{fig:rework}}
\end{figure}

If a faulty chip is detected on a bare module 
it is not further assembled. However, at this stage of the module assembly 
it is still possible to replace chips. 

To remove a readout chip the bare module is placed on a heatable chuck 
and held by vacuum as shown in Fig.~\ref{fig:rework}. The removal 
tool is then lowered onto the faulty readout chip and is also heated.
When a sufficient temperature is reached the chip is grasped by vacuum
and the tool is lifted. 
The heating temperatures and the lifting speed have to be adjusted in
a way that the indium balls stick to the sensor and not to the 
readout chip.

After this procedure a new readout chip can be placed in the
vacant position and the reflow has to be repeated. The bump yield
of replaced chips will not reach the level of the initial ones and
therefore all measures to minimize the fraction of bare modules 
to be reworked are taken. A careful inspection and cleaning of 
the readout chips as single dies is essential. The introduction of 
a dry cleaning step using a very soft sponge just before 
the flip chip procedure has reduced the number of chip failures 
drastically.

The completely tested bare module is then glued onto a base plate.
At this point a repair is not possible anymore. A complete description
of the module assembly is given in \cite{stefan}.

\section{Bump yield}

\begin{figure}
\centering\includegraphics[width=1\linewidth]{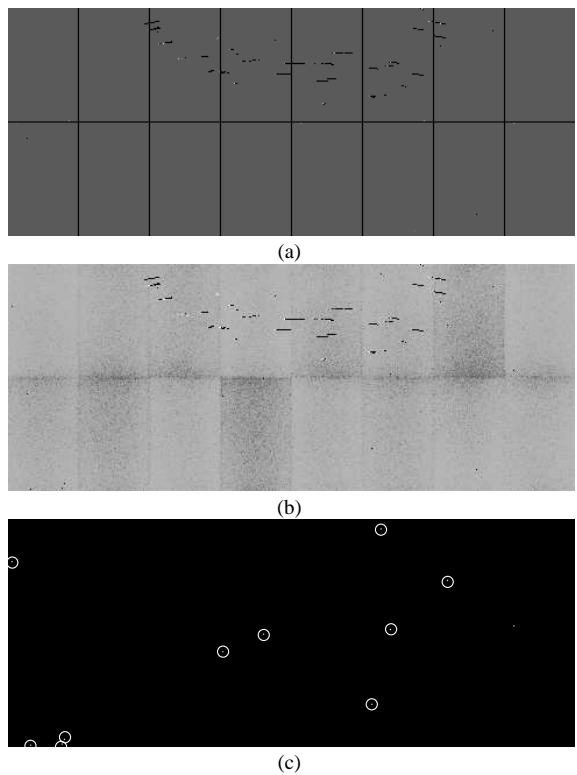}
\caption{Determination of the bump yield using
	illumination with X-rays. (a) Determination of
	broken channels in the readout chips, (b) illumination
	with X-rays, (c) missing bump bonds are marked with
	circles\label{fig:bump-yield}}
\end{figure}

The bump yield of a module can be measured either directly with
particles and X-rays or indirectly using the influence of the 
bump bond connection on the analogue behavior of the readout chip
as discussed in Ref.~\cite{andrey}. 

The bump yield determination of a PILATUS module 
with X-rays is illustrated in Fig.~\ref{fig:bump-yield}.
As a first step the electronics channels not responding to calibration
pulses are detected. The black pixels following a circular shape
in the upper middle of Fig.~\ref{fig:bump-yield}~(a) probably
suffer from a localized high sensor current leading to 
a saturation of the pixels' preamplifiers. These pixels are 
excluded from further analysis. As the second step the 
response of the module to X-rays is measured as shown in
Fig.~\ref{fig:bump-yield}~(b). The number of dead channels is
determined by looking for pixels which do respond to calibration pulses
but not to X-rays. In this case ten pixels (see Fig.~\ref{fig:bump-yield}~(c)) 
were found representing a fraction of less than $0.05\,\%$.

The experience gained with the 
production of a few tens of modules and some hundreds of readout chips
shows that, if no problems in the photolithographic steps occur,
the bump yield is either above 99.9\,\% or the bump bonding failes
completely (bump yield $< 90\,$\%). The occurrence of such fails
is rare and indicates a fundamental problem in the assembly procedure of
this particular module.

\section{Conclusions}

A bump bonding process was developed at PSI in the framework
of the R\&D for the CMS pixel detector. It features an underbump metalization 
composed of titanium, nickel and gold on both sensor and readout chip. The
indium bumps of about $20\,\umu$m height are deposited onto the sensor and
reflowed, while on the readout chip only a thin indium layer is evaporated.
After both parts are joined another reflow is performed to establish the
thermo-mechanical connection and to perform self alignment.

The bump yield of successfully built modules exceeds 99.9\,\%. The fraction of
such modules is not yet known, but the present experience is very encouraging
and there is hope to achieve a sufficient yield without reworking. 

The process was used to build the PILATUS~1M pixel detector with 18~modules 
\cite{gregor}.
The commissioning of a fully automatic chip placer increases the throughput 
considerably and the construction of large systems requiring 
hundreds of modules, like the CMS pixel detector, becomes feasible and will start in
early 2006.

\section{Acknowledgments}

Thanks are due to Stefan Ritter for the micrographs of the bumps 
(Fig.~\ref{fig:bump-foto}) and Beat Henrich for providing the 
bump yield measurement (Fig.~\ref{fig:bump-yield}).

\bibliographystyle{elsart-num}
\bibliography{bib_rohe}

\end{document}